\documentclass{IOS-Book-Article}

\usepackage{mathptmx}
\usepackage{soul}\setuldepth{article}
\usepackage{graphicx}
\usepackage{caption}
\usepackage{subcaption}
\usepackage{comment}
\usepackage{booktabs}
\usepackage{algorithm}
\usepackage{algpseudocode}
\usepackage{amsmath}
\usepackage{algorithm}
\usepackage{algpseudocode}
\usepackage{mathtools}
\usepackage{multicol}
\usepackage[framemethod=tikz]{mdframed}

\usepackage[show]{chato-notes}
\usepackage{paralist}
\usepackage{adjustbox}

\begin{document}

\pagestyle{headings}
\def\thepage{}
\begin{frontmatter}              % The preamble begins here.
%%%%%% Title %%%%%%%

%\pretitle{Pretitle}
%\title{Human Response to AI-Supported \mbox{Decision-Making in Face Matching:} \mbox{The Influence of Task Difficulty} and \mbox{Machine Accuracy}}
\title{Human Response to \mbox{Decision Support} in \mbox{Face Matching:} \mbox{The Influence} of \mbox{Task Difficulty} and \mbox{Machine Accuracy}}
%\subtitle{Subtitle}

%%%%%% Authors and affiliations %%%%%%%

%\author[]{Anonymous}
\author[A]{\fnms{Marina} \snm{Estévez-Almenzar}\orcid{0009-0005-8813-8593}%
\thanks{Corresponding Author: Marina Estévez-Almenzar, e-mail: marina.estevez@upf.edu.}},
\author[A]{\fnms{Ricardo} \snm{Baeza-Yates}\orcid{0000-0003-3208-9778}}
\\ and
\author[A,B]{\fnms{Carlos} \snm{Castillo}\orcid{0000-0003-4544-0416}}

\address[A]{Universitat Pompeu Fabra}
\address[B]{ICREA}

%%%%%% Abstract and keywords %%%%%%%
\begin{abstract}
% Context
Decision support systems enhanced by Artificial Intelligence (AI) are increasingly being used in high-stakes scenarios where errors or biased outcomes can have significant consequences.
% Questions
In this work, we explore the conditions under which AI-based decision support systems affect the decision accuracy of humans involved in face matching tasks.
% Previous work
Previous work suggests that this largely depends on various factors, such as the specific nature of the task and how users perceive the quality of the decision support, among others.
% Methods
Hence, we conduct extensive experiments to examine how both task difficulty and the precision of the system influence human outcomes.
%%%
% Findings
Our results show a strong influence of task difficulty, which not only makes humans less precise but also less capable of determining whether the decision support system is yielding accurate suggestions or not.
% Implications
This has implications for the design of decision support systems, and calls for a careful examination of the context in which they are deployed and on how they are perceived by users.
\end{abstract}


\end{frontmatter}

%%%%%% Header %%%%%%%
\def\hb{\hbox to 11.5 cm{}}
%\markboth{M. Estévez-Almenzar {\it et al.}}{June 2025 \hb}

% Comment this
%\thispagestyle{empty}
%\pagestyle{empty}

\section{Introduction}

Decision support systems are a key modality of use of Artificial Intelligence (AI).
They can be categorized by the extent to which the final decisions depend on them, from having no influence whatsoever, to being fully autonomous, with most cases operating somewhere between these two extremes~\cite{mcgee1998future,cummings2017automation}.
Ideally, these systems together with human operator(s) create a hybrid human-machine intelligence that exploits the expertise of human operators with the capacity to find patterns in historical data that yield better decisions.
This is particularly critical in applications that have significant effects on people, such as those described as \emph{high risk} by the European AI Act (EU Regulation 2024/1689) \cite{EUAIAct}. 
In these applications, given the ethical and legal requirements for human oversight, it is unlikely that fully automated systems are deployed in the near future.
Instead, hybrid systems combining human and machine intelligence are likely to become the norm. When humans and machines work together, they should be evaluated together \cite{matias2023humans}.
However, human-algorithmic behavior involves complex emerging patterns, uncertainties, and a certain degree of unpredictability that is in tension with the goal of developing safe and trustworthy systems.
Modeling how human operators respond to recommendations produced by an algorithm is paramount.

This is an active research topic, and previous work (surveyed in \S\ref{sec:related-work}) investigates aspects such as
the nature of the task for which decision support is provided, 
the way in which machine assistance is framed,
preconceptions that make users averse or over-reliant on algorithms, and the accuracy or perceived accuracy of the algorithmic suggestions, among other factors.
In this work, we study a face matching scenario in which a person is tasked with determining whether two photos correspond to the same person, and uses a face matching system as a decision support tool.
Our work addresses a series of research questions (\S\ref{sec:research-questions}) related to the extent to which a decision support system can enhance human accuracy, focusing on responses to both correct and incorrect suggestions and the effects of fluctuating (increasing or decreasing) system accuracy. We tackle these questions through a series of experiments (\S\ref{sec:experimental-setup}) conducted via crowdsourcing in which experimental variables include task difficulty, decision support accuracy, and whether a notification is given to users when decision support accuracy might change.
%
%We measure the accuracy of the final answer of users, compare it with the accuracy of their initial answer, and also ask them to what extent they considered the decision support useful.
Our results and discussion (\S\ref{sec:results}-\S\ref{sec:discussion}) show a strong influence of task difficulty, which not only makes human annotators less accurate, but also makes them more prone to be misled by inaccurate decision support and less aware of the accuracy of different decision support systems.
We also find important differences on human perceptions when decision support errors appear at the beginning or end of the sequence of tasks, compared to when they are randomly distributed throughout the sequence of tasks.
The last section (\S\ref{sec:conclusions}) presents our conclusions and outlines future work.

\section{Related Work}
\label{sec:related-work}

\subsection{Decision Support: The Nature of the Task}

Previous work highlights how the nature of the task plays a crucial role in human-machine interaction with a decision support system. 
One of the main axes of analysis is the distinction between objective and subjective tasks.
It has been shown that the more objective the task is perceived to be, the more likely the human is to be influenced by the machine \cite{castelo2019task}.
However, we also find evidence that in some cases the distinction between objective and subjective tasks does not play an important role in how people are influenced by machine suggestions \cite{logg2019algorithm}.
This apparent contradiction seems to be reconciled in the work of Hou {\it et al.} \cite{hou2021expert}, where the authors suggest that the really influential factor is the machine competence perceived by the human. This, in turn, depends heavily on how the decision support agent is presented: different framings of the same agent can shape different human perceptions, leading to inconsistent outcomes.
In a similar vein, Mahmud {\it et al.} \cite{mahmud2022influences} highlight the moral nature and the complexity of the task at hand. People tend to move away from the machine when it comes to making moral decisions, such as those related to legal or medical issues \cite{bigman2018people,gogoll2018rage,bonnefon2024moral}. Also, people tend to reject the machine when it comes to tasks that do not require high computational skills \cite{onkal2009relative,papenmeier2022accurate}. Furthermore, it has been noted that when utilitarian results hold significant value, there is a preference for AI recommenders instead of human ones, whereas when hedonic aspects are prioritized, there tends to be a resistance to AI recommenders in favor of human decisions \cite{longoni2022artificial}.

\subsection{Behavioral Patterns in Human-Machine Interaction}

Several patterns of behavior that emerge from the interaction between a human and a machine have been extensively studied.
\emph{Algorithmic aversion} is defined as a negatively biased perception of algorithms that manifests itself in a behavior of rejection towards the algorithm with respect to human agents. This aversion is especially reinforced when the human interacting with the machine observes that the machine makes mistakes \cite{dietvorst2015algorithm}. In contrast, in the evaluation of hybrid resolutions of moral problems humans tend to be evaluated more leniently than the machine, which is known as a human self-interest bias \cite{dong2024responsibility}.
%This could be closely related to the human self-interest bias observed in the evaluation of hybrid resolutions of moral problems, where the human agent is evaluated more leniently than the machine \cite{dong2024responsibility}.
%
Conversely, \emph{algorithmic appreciation} is known as a positive dominance of the algorithm that helps users avoid mistakes. However, this dominance might be detrimental if humans become overly dependent on machine outcomes, which may lead to errors \cite{cabitza2023ai}, that in some cases can have harmful consequences \cite{nonhumanerrors}. Another cognitive bias, well-recognized in psychology but less explored within the realm of human-machine interaction, is the tendency of humans to prefer information that aligns with their existing beliefs \cite{bashkirova2024confirmation}.

\subsection{Human-Machine Interaction in Facial Recognition}

Prior decisions made by face recognition systems influenced subsequent face matching decisions made by human operators. When face pairs were incorrectly labeled by the machine, the precision of humans decreased by drawing attention away from face images, even when humans were warned that machine predictions could be inaccurate \cite{fysh2018human}. Furthermore, decision-deferral rates in human-machine systems influence both human performance and trust during face-matching tasks \cite{salehi2021decision}. In addition, the type of errors done by humans and systems are different \cite{facialrecognitionerrors}.

It is well documented that human face recognition accuracy can be improved by the wisdom of crowds: combined judgment of many is better than the decision of an individual \cite{jeckeln2018wisdom}. There is a similar benefit to merging the performance of multiple algorithms \cite{ranjan2019fast}. When considering decisions resulting from the fusion of human decision and machine decision, the results can lead to large performance improvements compared to the human response or the algorithm response alone \cite{phillips2018face}.

\smallskip
Our work addresses task complexity, a subject that has not been explored in as much detail as other aspects in the surveyed literature \cite{salimzadeh2023missing}, positioning it as a pivotal element to be carefully considered in the design of decision support systems.
%
%For the purposes of our study, we define complexity as the difficulty perceived by the human agent, thus prioritizing the human factor in a field that may involve high risks.
%
Additionally, in many real-world scenarios in which data evolves, and given that machine learning models should be trained and applied on data with identical distributions, keeping models up to date is a critical task \cite{majidi2024efficient,faubel2023towards,bayram2024towards}.
Updating the models produces variability in performance, causing an effect on interaction patterns \cite{renier2021err}. 
As far as we know, previous research has not thoroughly examined the effects of potential machine variability on human-machine interactions, nor has it been thoroughly studied the effects of how errors are distributed along a sequence of tasks in machines of equal average accuracy. In this paper we consider machines of varying accuracy and observe how these variations affect human performance. We also study whether notifying the human operator each time a variation in the machine occurs makes any difference in joint human-machine performance. %This machine variability allows us to compare, for machines with the same average accuracy, how it affects the way errors are distributed in the interaction flow. 

%
%Moreover, this paper boldly tackles the intricate notion of task complexity, a subject scarcely explored in the existing literature \cite{salimzadeh2023missing}, positioning it as a pivotal component to be meticulously considered in the strategic design of decision support systems. For the purposes of our study, we define complexity through the lens of the challenges and obstacles perceived by the human agent, thereby enhancing our understanding of the human-machine interaction dynamic.

%We compare the effect of these variable machines with that of three different but static precision machines. To avoid biasing the participant's perception of the machine \cite{hou2021expert}, we chose to introduce all machines as neutrally as possible, without specifying whether one machine is better than another. This will allow us to study more concretely the perception generated by the participant themselves, without conditioning them. 
%
%As to the nature of the face-matching task, we consider that it is hardly categorized as objective or subjective: given a pair of faces, identifying if they are the same person is subjective in that there may be a wide range of human different opinions, but at the same time there is an unquestionable ground truth concerning the identity of the persons being identified. Whether this task has a moral component, or whether it can be categorised as hedonistic or utilitarian, is also debatable, as it depends on the domain of application of the face matching task. 

\section{Research Questions}
\label{sec:research-questions}

Our experiments are designed to address the following research questions:

\textbf{RQ1} \textit{Does an AI-based decision support system improve human performance in a face matching task?} We test to what extent the support of a high-accuracy machine improves human accuracy in solving a face-matching task, while testing whether this improvement depends on the difficulty of the task.

\textbf{RQ2} \textit{Does a low accuracy AI decision support system improve or deteriorate human performance in a face matching task?}\newline
We test to what extent the support of a low-accuracy machine deteriorates human accuracy, while testing whether this deterioration depends on the difficulty of the task.

\textbf{RQ3} \textit{Does a variable accuracy AI decision support system improve or deteriorate human performance in a face matching task?}\newline
We test to what extent the support of a variable-accuracy machine improves or deteriorates human accuracy, while testing whether this change depends on the difficulty of the task. We also test whether this change depends on human awareness of this machine variability.

\section{Experimental Setup}
\label{sec:experimental-setup}

To investigate our research questions, we designed three experiments. 
We considered three independent variables: \begin{inparaenum}[(i)]
\item problem difficulty, 
\item machine accuracy, and 
\item change notifications.
\end{inparaenum}
\emph{Problem difficulty}, described in \S\ref{sec:procedure}, indicates the difficulty of the tasks assigned to the participant. 
\emph{Machine accuracy}, described in \S\ref{sec:machines}, indicates the accuracy of the decision support assigned to the participant.
\emph{Change notification} indicates whether the participant is notified or not when the decision support system changes, and only applies to the experiment in which the accuracy varies.
We measured three dependent variables:
accuracy, influence factor, and confirmation factor. The three dependent variables are defined in \S\ref{sec:measurements}.

\paragraph{Experiment 1: With / Without Decision Support}
This experiment works as our ``control experiment'' and is designed for the purpose of investigating RQ1. Two groups of different participants were compared.
The control group solved the task without machine suggestions.
The experimental group of participants received, for every task, a suggestion from a system having 95\% accuracy.

\paragraph{Experiment 2: With Degraded Decision Support}
This experiment is designed for the purpose of investigating RQ2. Two experimental groups were compared.
One of them solved the face matching tasks while receiving suggestions from a 5\% accuracy machine and the other one while receiving suggestions from a 50\% accuracy machine.

\paragraph{Experiment 3: With Variable Decision Support Machine} This experiment is designed for the purpose of investigating RQ3. Four groups of different participants were compared. Two groups solved the tasks while receiving suggestions from an \textit{increasing} accuracy machine (that we note INC machine). One of these groups was notified every time the machine changed, and the other group was not.
Conversely, two other groups solved the tasks while receiving suggestions from a \textit{decreasing} accuracy machine (DEC), and similarly, one of them was notified of changes, while the other was not.
%Finally, the other two groups solved the tasks while receiving suggestions from a \textit{temporarily failed} accuracy machine (that we note FAIL machine), one receiving the notification and the other not.

Our research plan was reviewed and approved by %the Institutional Committee for Ethical Review of Projects (CIREP) at Universitat Pompeu Fabra.
the Ethics Review Board of the university of the lead author. % ANONYMIZED
The review included compliance with internationally accepted ethical principles in research, and with personal data protection, in particular by the EU General Data Protection Regulation (2016/679).

\begin{comment}
\begin{table}[]
\centering
    \caption{Summary of experiments. Each experiment was carried out twice, once for each problem difficulty (``normal'' or ``hard'').}
    \label{tab:exp_setup}
    \begin{tabular}{@{}lcc|cc|cccc@{}}
    \toprule
    & \multicolumn{2}{c|}{\textit{Experiment 1}} & \multicolumn{2}{c|}{\textit{Experiment 2}} & \multicolumn{4}{c}{\textit{Experiment 3}}                                      \\ \midrule
    \textit{Machine accuracy} & none  & 95\%  & 5\%  & 50\% & \multicolumn{2}{c}{INC} & \multicolumn{2}{c}{DEC} \\
    \textit{Change notification} & -  & -  & -  & -  & yes  & no  & yes  & no  \\
    \# participants & 20  & 20  & 20  & 20   & 20  & 20  & 20 & 20 \\ \bottomrule
    \end{tabular}
\end{table}
\end{comment}

\subsection{Procedure}
\label{sec:procedure}

We performed an online user study, with the following structure. 

\paragraph{Participant recruitment}
We recruited participants through a crowdsourcing platform for experimentation named Prolific.\footnote{\href{https://www.prolific.co/}{www.prolific.co}}
We considered four countries in continental Europe in which Prolific has large user bases: France, Germany, Italy, and Spain, plus the United Kingdom. 
We made sure that our sets of participants were gender balanced, a feature that the platform provides based on participants' disclosures of gender.
In total, we recruited 320 participants, and each participant annotated 30 different pairs of images. In total, we collected 9,600 participant's annotations, and a total of 60 different pairs were annotated under various conditions.
Participants were paid 9.16 EUR per hour. Specifically, they were paid 1,07 EUR for labeling 30 pairs of images, with an average completion time of 10 minutes.
To encourage participants' effort, we set a bonus: participants were warned (and reminded during the study) that if they managed to correctly match more than 80\% of the pairs (more than 24 pairs) they would receive an extra payment of 30\%.

\paragraph{Tasks selection}
%\mynote[from=ChaTo]{You can reference your previous paper here, assuming there is a pre-print somewhere, or if not cite as "forthcoming". However, if the conference requires anonymization, this needs to be done with care.}
Each of the experiments was carried out twice. Once with a set of pairs that we call \textit{Normal Set}, and once with a different set of pairs that we call \textit{Hard Set}. Both sets consist of pairs from the DemogPairs \cite{hupont2019demogpairs} testing database, and their classification as \textit{Normal} or \textit{Hard} is based on a preliminary study in which a total of 540 pairs were tested by a total of 162 participants (10 different pairs per participant, 3 different participants per pair). These participants were paid the same as described previously. In this study, participants rated these pairs by answering the question ‘Are they the same person?’ with the options ``No'', ``Probably not'', ``Not sure'', ``Probably yes'', or ``Yes''. This allows us to categorize some of these pairs according to the difficulty experienced by participants in solving the task. In this previous study, these 540 pairs were also evaluated by two state-of-the-art models jointly. These two models are IR50+ArcFace and LightCNN, and the joint accuracy over DemogPairs test dataset is above 95\%. By ``joint accuracy'' we mean the accuracy based on the mean response of both models. This ensemble machine is the base decision support system we use in our experiments.
%From now on, we will call this joint machine the \textit{Original Machine}, as it will be the machine on which the simulation of the machines necessary for the experiments of this work will be based. \textit{Hard Set} and \textit{Normal Set} are subsets of the set of pairs that were correctly classified by the \textit{Original Machine}.

Hard Set: We select those pairs whose mean participant response was very close to ``Not sure'', and at least one of the three participants made a mistake. We also selected those pairs where the mean response is exactly ``Not sure''. We obtained a total of 69 pairs (63 positive and 6 negative). After a careful manual inspection, from among the 63 positive pairs we chose the 24 most difficult ones which together with the 6 negative ones (with no occlusion {\it i.e.}, no objects obstructing parts of the face) form the \textit{Hard Set}.

Normal Set: We define another set of pairs with a slightly more relaxed human certainty condition: we select those pairs whose mean response is close to ``Not sure'', avoiding those whose mean is exactly ``Not sure''. From these pairs we randomly choose 30 pairs of images, maintaining the above ratio of 24 positive and 6 negative pairs, and avoiding repetitions with the ``Hard Set''. This left us with 30 pairs that will form the \textit{Normal Set}.

\begin{figure}[t!]
     \centering
     \begin{subfigure}[b]{0.49\textwidth}
         \centering
         \includegraphics[width=\textwidth]{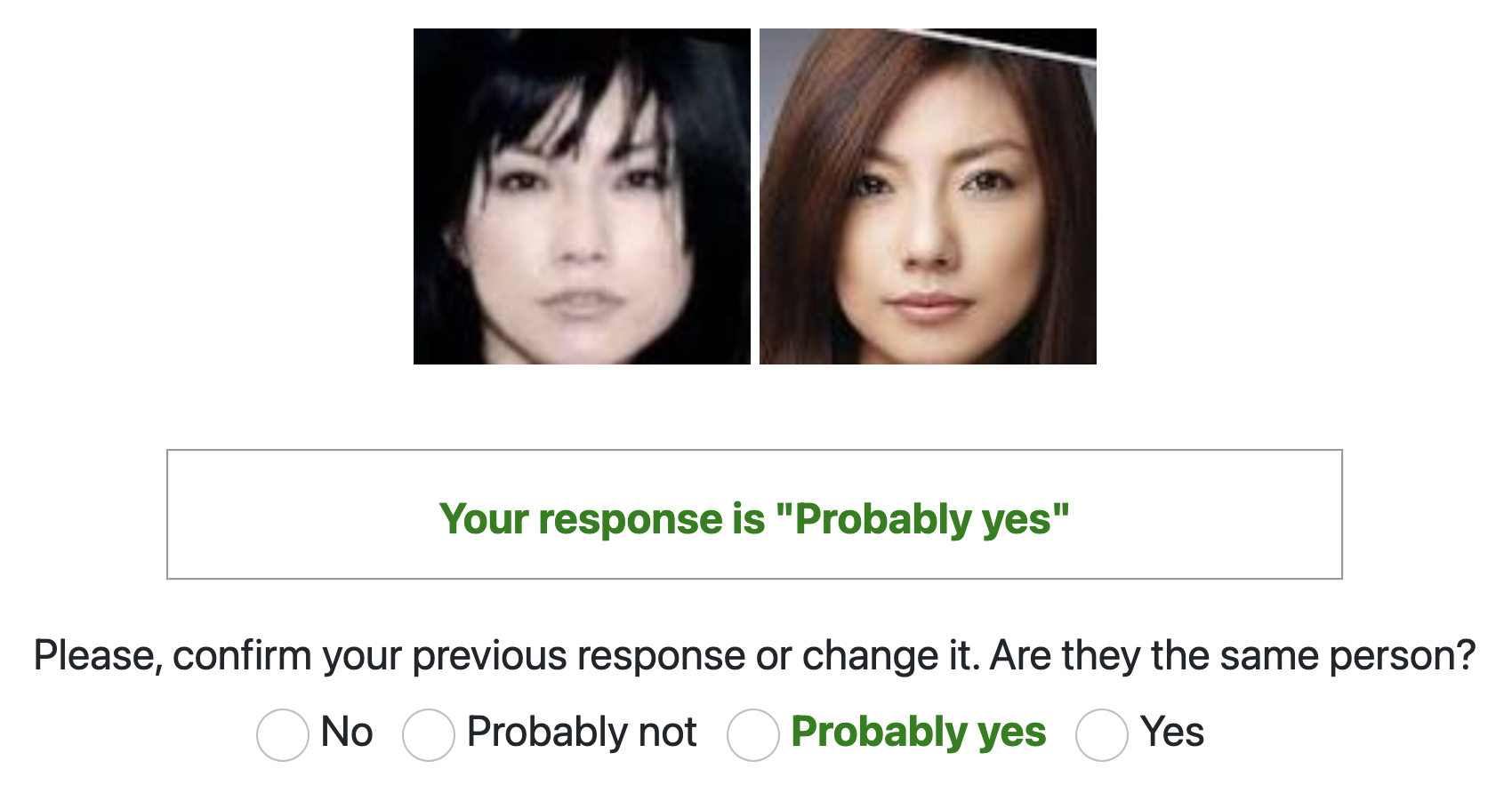}
         \caption{Control group}
         \label{fig:survey_2_control}
     \end{subfigure}
     \begin{subfigure}[b]{0.49\textwidth}
         \centering
         \includegraphics[width=\textwidth]{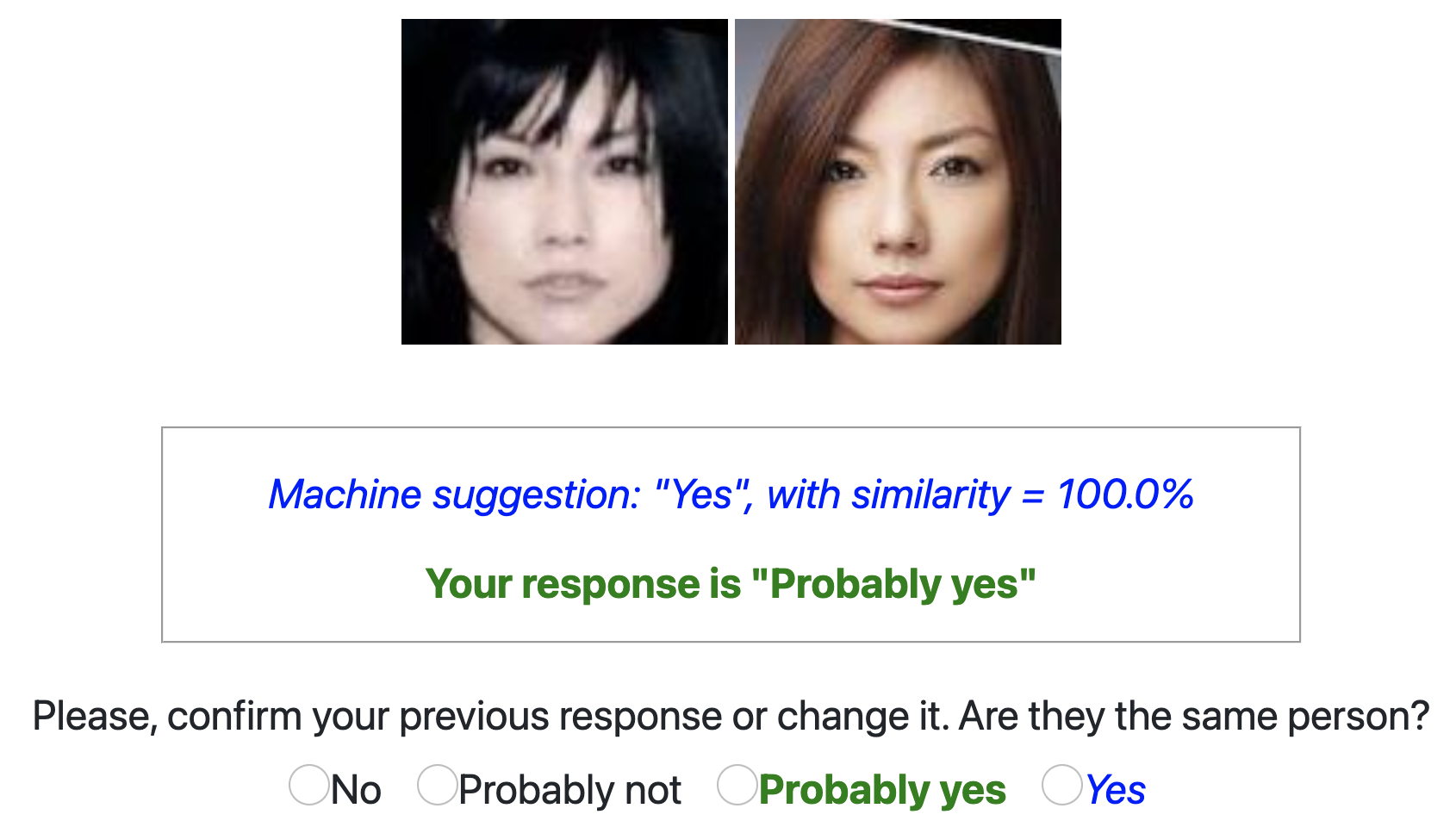}
         \caption{Experimental groups}
         \label{fig:survey_2_machine}
     \end{subfigure}
     \caption{Survey screenshots.}
\end{figure}

\paragraph{Face matching tasks}
Participants evaluated one pair of images at a time. Given a pair $p$, the participant had to answer the question \textit{Are they the same person?}, with the possible options: \textit{No}, \textit{Probably not}, \textit{Probably yes}, or \textit{Yes}. 
%(numerically represented as -2, -1, 1, 2, respectively). Note that the distance between the numerical representation of each answer is consistent with its logic: \textit{Probably not} and \textit{Probably yes} are further away than \{\textit{No}/\textit{Yes}\} and \{\textit{Probably not}/\textit{Probably yes}\} respectively, since the former represents a complete change of opinion.
After answering, if the participant was assigned to one of the experimental groups, the machine suggestion was shown together with the machine similarity score associated with the pair $s_p$ on which the suggestion was based: $0\leq s_p\leq 0.25$ with suggestion \textit{No}, $0.25< s_p\leq 0.50$ with suggestion \textit{Probably not}, $0.50< s_p\leq 0.75$ with suggestion \textit{Probably yes} or $0.75< s_p\leq 1.00$ with suggestion \textit{Yes}. They had the possibility to modify their answer (see Figure \ref{fig:survey_2_machine}). Participants in the control experiment also had the possibility to modify their answer (see Figure \ref{fig:survey_2_control}).

\begin{comment}
\begin{figure}
     \centering
     \begin{subfigure}[b]{0.49\textwidth}
         \centering
         \includegraphics[width=\textwidth]{images/survey_q1.png}
         \label{fig:survey_q1}
     \end{subfigure}
     \begin{subfigure}[b]{0.49\textwidth}
         \centering
         \includegraphics[width=\textwidth]{images/survey_q2.png}
         \label{fig:survey_q2}
     \end{subfigure}
     \caption{Questionnaire screenshots.}
\end{figure}
\end{comment}

\paragraph{Exit survey}
After evaluating the 30 pairs, the participants who interacted with a machine completed an exit survey. They were asked whether the suggestions of the machine had been useful to 1. \textit{make a decision when they were not sure about their answer}, 2. \textit{make a decision when some of the images were blurry or the quality was not good}, 3. \textit{make decisions faster}, 4. \textit{confirm their decision when they were certain about it}, 5. \textit{make more accurate decisions}, and 6. \textit{make them feel more confident about the answer}. Participants had the possibility to answer \textit{Strongly disagree} / \textit{Disagree} / \textit{Neither agree nor disagree} / \textit{Agree} / \textit{Strongly agree}.

\subsection{Machines}
\label{sec:machines}
%The performance of the joint machine is based on giving an answer based on the mean of the two similarity scores obtained by each model (IR50+ArcFace and LightCNN). 
To simulate machines that adapt to the circumstances that we want to reproduce in each experiment, we introduce noise into the ensemble machine. Given a pair $p$, we obtain the associated similarity score $s_p$, with $0\leq s_p \leq1$. We define the noise as $f(s_p) = 1 - s_p$, which gives us a noisy similarity score that forces the opposite response. The number of pairs to which this noise is applied depends on the machine to be simulated. 
These machines are described below.
\paragraph{Static machines} We simulate three machines: 95\%, 50\%, and 5\% accuracy machines. Each of them is a realization of the probabilistic situations we simulate, which means that they have exactly the target accuracy in every experiment. 
\paragraph{Variable machines}
We simulate two variable accuracy machines: 1) \textit{INC machine} simulates a model that increases its accuracy over time, and is defined as the concatenation of three machines: 5\% - 50\% - 95\% accuracy machines, each of them solving one third of the total number of tasks. 2) \textit{DEC Machine} simulates a model that decreases its accuracy over time, and is defined as the concatenation of three machines: 95\% - 50\% - 5\% accuracy machines, each of them solving one third of the total number of tasks.
%\paragraph{FAIL Machine} With this machine we simulate a model that temporarily suffers an error. Given a set of pairs, this machine is defined as the concatenation of three machines: a 95\% Machine for the first third of the tasks, a 5\% Machine for the second third, and a 95\% Machine for the last third of the tasks.

Regarding change notification, there were two conditions: without notification, and with notification.
Participants in the first condition were not told anything about variable machine accuracy.
Participants in the notification condition, before starting the survey, were shown the following message:  \textit{There are three AIs, named machine A, machine B, and machine C. We will notify you every time there is a change}.
The notification message was: \textit{Next, you will receive suggestions from machine \{A / B / C\}}.

\subsection{Measurements}
\label{sec:measurements}

\paragraph{Participant accuracy} We compute the fraction of correct responses, with respect to the ground truth, given by the participant before (\textit{initial accuracy}) and after (\textit{final accuracy}) seeing the machine suggestion.
In the analysis we show the macro-average across participants of the initial and final accuracy, together with their standard deviation.

\paragraph{Interaction Factors}
Given a pair $p$ solved by a participant, let $r_i$ be the \textit{participant's initial response}, $r_f$ the \textit{participant's final response}, and $m$ the \textit{machine suggestion}.\footnote{$r_i,r_f\in\{-2,-1,1,2\}\equiv\text{\{\textit{No, Probably not, Probably yes, Yes}\}}, m = -2+4s_p\in[-2,2] \text{ with }s_p\in[0,1]$} 
%We considered two human-machine interaction factors:
% with $r_i,r_f\in\{-2,-1,1,2\}$, $s_p\equiv$\textit{ similarity score associated to p}, with $s_p\in[0,1]$, $m\equiv$\textit{ machine suggestion}, $m = -2+4s_p\in[-2,2]$

{\em Influence Factor}: We measured how much the machine suggestion influenced the participant's response. This value, that we denote by $IF$, ranges from -1 to 1, and is based on the influence measure defined by Hou {\it et al.} in \cite{hou2021expert}. 
We define
    \begin{equation*}
    IF(r_i,m,r_f) =
        \begin{cases*}
          -|r_f-r_i| & if $|m-r_i|<1$ \\
          \frac{r_f-r_i}{m-r_i} & if $|m-r_i|\geq1$
        \end{cases*}
    \end{equation*}
In the analysis we show the macro-average of the influence factor. We also measure the probability that this influence is positive, $P(IF>0)$, zero $P(IF=0)$, or negative $P(IF<0)$. We find this metric easy to interpret as, for example, a positive value is given to cases where the participant changes their response in the direction of what is recommended by the machine. %and the probability that it is equal to zero, $P(IF==0)$. 

{\em Confirmation Probability}: We measured the probability that the participant's initial response and the machine's suggestion match and the participant does not change their final response. This event, that we note as $C$, occurs when $r_i = r_f$ and $|m-r_i|\leq 1$. 
%In this case, the influence factor will be zero, not because there was no influence, but because the human and machine responses coincided. 
In the analysis we show the probability that this occurs, that we note as $P(C)$.

\begin{table}[t!]
\caption{Average participant initial accuracy $a_i$, final accuracy $a_f$, difference among both $\delta$, influence factor $IF$, probability that this influence is positive $P(IF>0)$, probability that this influence is neutral $P(IF==0)$, probability that this influence is negative $P(IF<0)$, and probability of confirmation $P(C)$ for all the experiments with the \textit{Normal Set} and the \textit{Hard Set}. There are 20 participants for every row. (n) stands for \textit{with notification}, (-) for \textit{with no notification}.}
\label{tab:normal_and_hard}
\begin{adjustbox}{width=0.90\textwidth}
\begin{tabular}{@{}lccc|ccccc@{}}
\toprule
%& \multicolumn{5}{c}{Normal set} \\ \midrule
\textit{Normal Set}   &  $a_i$ & $a_f$   & $\delta$ &     $IF$ &  $P(IF>0)$ & $P(IF=0)$ & $P(IF<0)$  &     $P(C)$\\ \midrule
no machine        &  0.67 ± 0.23 & 0.67 ± 0.23 & 0     & -  & - & - & -  & -   \\
95\%     &  0.71 ± 0.20 & 0.80 ± 0.20 & +0.09 & 0.01 & 0.13 & 0.78 & 0.09 & 0.57   \\ \midrule
50\%     & 0.69 ± 0.21 & 0.66 ± 0.21 & -0.03 & 0.05 & 0.14 & 0.81 & 0.05 & 0.40  \\
5\%      & 0.64 ± 0.28 & 0.61 ± 0.30 & -0.03 & -0.03 & 0.07 & 0.87 & 0.06 & 0.32 \\ \midrule
INC (n)  & 0.64 ± 0.21 & 0.64 ± 0.22 & 0     & +0.00 & 0.11 & 0.82 & 0.07 & 0.45 \\ 
INC (-)  & 0.64 ± 0.21 & 0.63 ± 0.18 & -0.01 & +0.00 & 0.12 & 0.79 & 0.09 & 0.47 \\
DEC (n)  & 0.70 ± 0.21 & 0.69 ± 0.21 & -0.01 & 0.04  & 0.11 & 0.85 & 0.04 & 0.42 \\
DEC (-)  & 0.71 ± 0.18 & 0.67 ± 0.21 & -0.04 & 0.10  & 0.22 & 0.70 & 0.08 & 0.39 \\ \bottomrule
\end{tabular}
\end{adjustbox}
\begin{adjustbox}{width=0.90\textwidth}
\begin{tabular}{@{}lccc|ccccc@{}}
\toprule
%& \multicolumn{5}{c}{Hard set}                  \\ \midrule
\textit{Hard Set}   &  $a_i$ & $a_f$   & $\delta$ &     $IF$ &  $P(IF>0)$ & $P(IF=0)$ & $P(IF<0)$   &     $P(C)$\\ \midrule
no machine        &  0.57 ± 0.20 & 0.57 ± 0.20 & 0& - &- & - & -  & -    \\
95\%     &  0.58 ± 0.23 & 0.65 ± 0.22 & +0.07 & 0.13 & 0.15 & 0.81 & 0.04 & 0.34 \\ \midrule
50\%     & 0.55 ± 0.24 & 0.56 ± 0.22 & +0.01 & 0.24 & 0.23 & 0.75 & 0.02 & 0.34 \\
5\%      & 0.54 ± 0.22 & 0.38 ± 0.24 & -0.16 & 0.21 & 0.22 & 0.74 & 0.04 & 0.26 \\ \midrule
INC (n)  & 0.55 ± 0.15 & 0.56 ± 0.15 & +0.01 & 0.18 & 0.18 & 0.78 & 0.04 & 0.35 \\ 
INC (-)  & 0.49 ± 0.17 & 0.49 ± 0.17 & 0     & 0.17 & 0.18 & 0.80 & 0.02 & 0.35 \\
DEC (n)  & 0.58 ± 0.17 & 0.59 ± 0.17 & +0.01 & 0.11 & 0.15 & 0.80 & 0.05 & 0.36 \\
DEC (-)  & 0.53 ± 0.16 & 0.52 ± 0.18 & -0.01 & 0.13 & 0.19 & 0.72 & 0.09 & 0.30 \\ \bottomrule
\end{tabular}
\end{adjustbox}
\end{table}

\newpage

\section{Results}
\label{sec:results}

Table~\ref{tab:normal_and_hard} summarizes our experimental results, which we detail and explain next.

\subsection{Experiment 1: With vs. Without Decision Support Machine}
For every set, we compare two groups of participants: those who did not receive machine suggestions and those who received suggestions from the 95\% accuracy machine. 
%We compare their accuracy, their influence factor and their confirmation factor.

\paragraph{Normal Set}
Almost all participants with no machine suggestions maintained their initial response when given the opportunity to modify it in the vast majority of tasks, making both the initial and final accuracy 0.67 ± 0.23.
Results from participants interacting with the 95\% accuracy machine suggest that the support of a high accurate machine improves human performance (see left plot in Figure \ref{fig:violins}) even when the influence is low.
%0.17, while the mean confirmation factor is 0.30.

\paragraph{Hard Set}
Similarly, almost all participants with no machine suggestions maintained their initial response when given the opportunity to modify it in the vast majority of tasks, making both the initial and final accuracy 0.57 ± 0.20.
As before, participants interacting with the 95\% accuracy machine got to improve their accuracy (see right plot in Figure \ref{fig:violins}). For these participants, the influence is higher than for those in the Normal Set. Participants perceive the 95\% accuracy machine more positively in the Normal Set, as shown in Figure \ref{fig:questionnaire}, despite its higher influence in the Hard Set.

\subsection{Experiment 2: With Degraded Decision Support Machine}
\begin{figure}[t]
    \centering
    \begin{subfigure}[b]{0.49\textwidth}
         \centering
         \includegraphics[width=\textwidth]{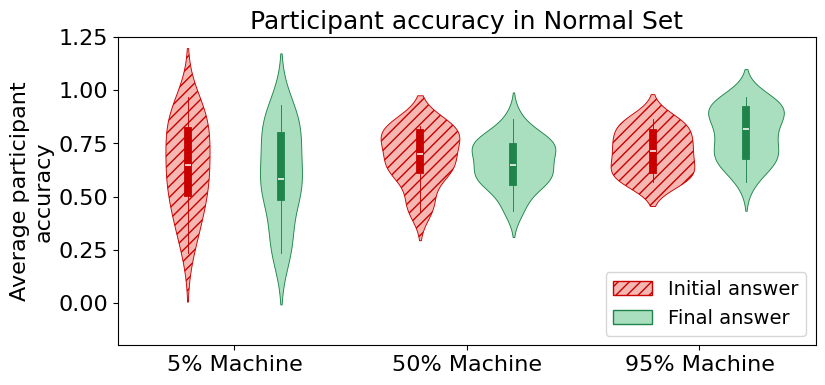}
         %\label{fig:violins_normalset}
     \end{subfigure}
     \begin{subfigure}[b]{0.49\textwidth}
         \centering
         \includegraphics[width=\textwidth]{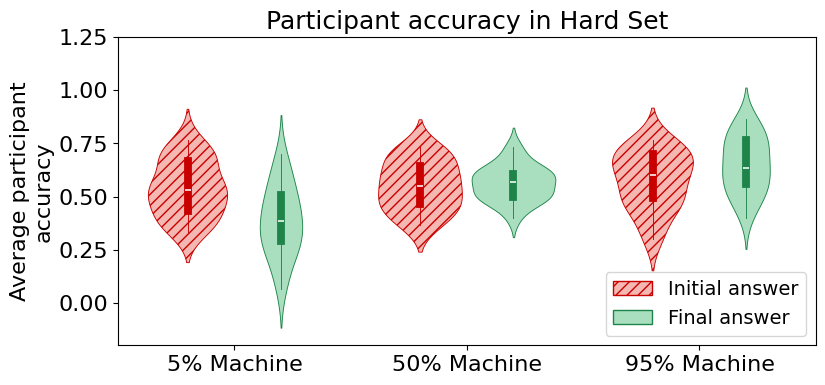}
         %\label{fig:violins_hardset}
     \end{subfigure}
    \caption{Participant initial and final average accuracy distributions for the \textit{Normal Set} and the \textit{Hard Set}, for 5\%, 50\%, and 95\% Machines in Experiment 1.}
    \label{fig:violins}
\end{figure}

In this experiment, we consider 5\% and 50\% accurate machines. %For every set of pairs, we compare the group of participants who received suggestions from the 5\% accuracy machine with those who received suggestions from the 50\% accuracy machine.
\paragraph{Normal Set}
Participants who interact with the 50\% accuracy machine and with the 5\% accuracy machine experience a drop in accuracy, suggesting that low accuracy support deteriorates human performance (see left plot in Figure \ref{fig:violins}). Participants on a 5\% accuracy machine showed a slightly negative influence (capacity for correcting mistakes from the machine).

\paragraph{Hard Set} 
For participants interacting with the 50\% precision machine, the accuracy does not vary markedly (see the right plot in Figure \ref{fig:violins}), although the influence is high.
For participants interacting with the 5\% accuracy machine, there is a marked deterioration in accuracy (see right plot in Figure \ref{fig:violins}). For these participants, the influence is also high and they tend to be misled by the machine more often than in the normal set.

\begin{figure}[t]
    \centering
    \includegraphics[width=0.8\linewidth]{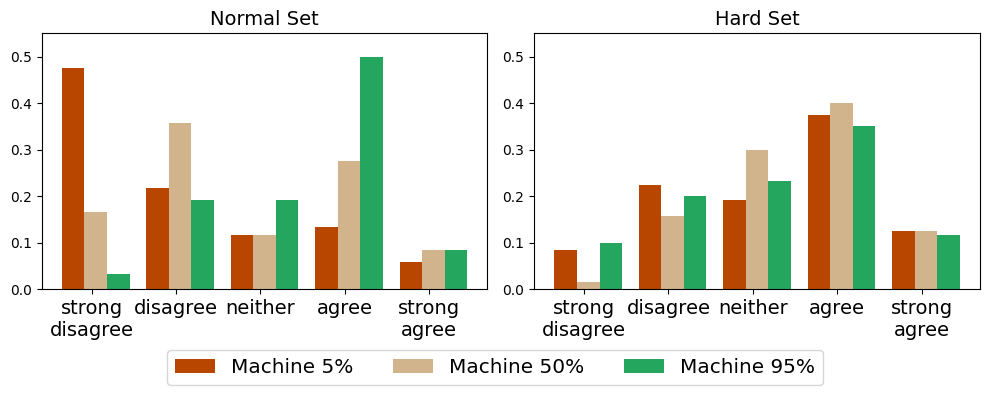}
    \caption{Results from the exit survey, from the participants who interacted with some static machine. They were asked whether the suggestions of the machine had been useful to \textit{make a decision} 1. \textit{when they were not sure}, 2. \textit{when some of the images were blurry or the quality was not good}, 3. \textit{faster}, to 4. \textit{confirm their decision when they were certain about it}, 5. \textit{make more accurate decisions}, and 6. \textit{make them feel more confident about the answer}. We show the average of the responses across the six questions.}
    \label{fig:questionnaire}
\end{figure}

In Figure \ref{fig:questionnaire}, we can see that for the participants who label pairs from the Normal Set, the more accurate the machine is, the more useful they find it. However, for the participants who label pairs from the Hard Set, the accuracy or inaccuracy of the machines does not affect as much the perceived usefulness of the machine. 
The Kolmogorov-Smirnov significance test reveals that for the 5\% accuracy and 50\% accuracy machines, there are significant differences between the distribution of perceived usefulness with the Normal Set and with the Hard Set (KS = $0.39$, $p \ll 0.0001$; KS = $0.35$, $p \ll 0.0001$, respectively). No significant difference is observed for the 95\% accuracy machine.

\subsection{Experiment 3: With the Variable Decision Support Machine}
\begin{figure}[t]
    \centering
    \includegraphics[width=1\linewidth]{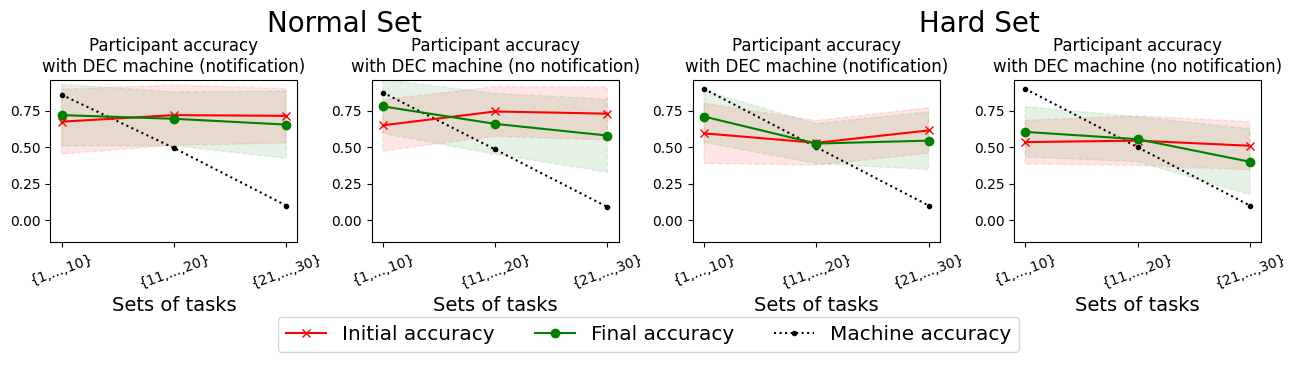}
    \caption{Participant accuracy for DEC machines (with and without notification) with the \textit{Normal Set} (left) and the \textit{Hard Set} (right).}
    \label{fig:exp_easy_and_hard_DEC}
\end{figure}

For every set of pairs, we compare the group of participants who received suggestions from \textit{INC Machine} (5\% \textrightarrow 50\% \textrightarrow 95\%) and \textit{DEC Machine} (95\% \textrightarrow 50\% \textrightarrow 5\%). We distinguish between those participants who were notified every time the machine changed and those who were not.
\paragraph{Normal Set} With the \textit{INC Machine}, for both participants with and without notification, accuracy does not vary markedly, and there is no discernible influence by any of the machines. In both cases, confirmation was around 50\%.
With the \textit{DEC Machine}, for participants with notification, accuracy does not vary markedly, and there is no discernible influence (see plots in first column in Figure \ref{fig:exp_easy_and_hard_DEC}).
For participants without notification, the results show a deterioration in the participant's accuracy (see plots in second column in Figure \ref{fig:exp_easy_and_hard_DEC}), and the influence factor noticeably exceeds the influence of participants with notification.

\paragraph{Hard Set} With both \textit{INC Machine} and \textit{DEC Machine}, for both participants with and without notification, accuracy does not vary markedly. In contrast with participants in the \textit{Normal Set}, the influence is now noticeable.

\begin{comment}
\begin{figure}
\centering
\includegraphics[width=1\linewidth]{images/exp2_all_acc_factors.png}
\caption{Participant accuracy, Influence and Confirmation factors for INC and DEC machines with \textit{Hard Set}}
\label{fig:exp2_all_acc_factors}
\end{figure}
\end{comment}

\section{Discussion}
\label{sec:discussion}

\begin{mdframed}[backgroundcolor=black!5]
\textbf{RQ1} \textit{Does an AI decision support system help improve human performance in a face matching task?}

An accurate machine may improve human performance, but the difficulty of the task might prevent the human from fully exploiting this advantage.
\end{mdframed}
A high accuracy machine improves human performance in both easy and hard tasks, which is aligned with previous works in the literature \cite{logg2019algorithm,araujo2020ai}. The influence of this machine is higher when the tasks are harder.
%, but more effective when the tasks are easy. This means that a low level of influence of an accurate machine on easy tasks is more beneficial than a high level of influence of an accurate machine on difficult tasks.
However, the improvement and high influence on the Hard Set do not appear to stem from the participant's capacity to recognize the machine's high accuracy since the final questionnaire indicates that inaccurate machines are viewed as equally useful.
\newpage
\begin{mdframed}[backgroundcolor=black!5]
\textbf{RQ2} \textit{Does a low accuracy AI decision support system improve or deteriorate human performance in a face matching task?}

High task difficulty allows an inaccurate machine to induce error more than an accurate machine can induce correctness, probably due to the participants' inability to really grasp how inaccurate the machine is.
\end{mdframed}
For degraded accuracy machines supporting human performance in easy tasks, the degradation of the participant's accuracy is hardly noticeable, suggesting that for a set of easy tasks the participant is able to solve without attending to the machine, as corroborated by the close-to-zero influence values. Observe that the minimal impact of the 95\% accuracy machine is partially due to the high confirmation rate ({\it i.e.}, the machine and user frequently agree, thus reducing the chance of influence), whereas the 5\% accuracy machine exhibits a lower confirmation rate yet maintains a marginal influence ({\it i.e.}, in many tasks, the machine has opposed the participant but did not alter their viewpoint). This suggests that for easy tasks, the participant knows how to solve it well, as it matches the high-accuracy machine, contradicting the low-accuracy one.

However, for difficult tasks, a very low accuracy machine can induce a participant to error very noticeably, even far exceeding the ability of a high-accuracy machine to induce correctness. Our results suggest that for difficult tasks participants tend to be influenced more by the low-accuracy machines than by the accurate machine, while the opposite is true for easy tasks.
This may stem from a \textit{projection} of the difficulty experienced by participants. Additionally, the control group's accuracy (no machine) aligns with the 50\% accuracy machine. A potential \textit{mirroring} between this machine and its users could explain its significant influence. These phenomena (\textit{self-projection} and \textit{mirroring}) are well established in psychology \cite{waytz2011two}, yet under-explored in human-machine interaction, to the best of our knowledge.
Moreover, results from the exit survey highlight that for difficult tasks the participant is not able to perceive a difference in the usefulness of interacting with very accurate or inaccurate machines, which is aligned with some research in the literature \cite{papenmeier2022accurate}.

\begin{mdframed}[backgroundcolor=black!5]
\textbf{RQ3} \textit{Does a variable AI decision support system improve or deteriorate human performance in a face matching task?}

Automation bias can be induced in a low-performing machine that initially provides accurate support. This can be mitigated with a simple notification that the machine has changed.
\end{mdframed}
For variable machines supporting human performance in easy tasks, the machine influence is barely noticeable except in one case: when a machine initially functions with high precision but gradually loses accuracy. If the participant remains unaware of any change in the machine, they continues to rely on the machine, leading to mistakes.
This aligns with the logic of some patterns observed in the literature, where algorithmic aversion is seen to increase when the user sees that the machine fails \cite{dietvorst2015algorithm}. We observe something analogous: participants can move from algorithmic appreciation to automation bias after observing machine success. This situation can be prevented by notifying the participant about a machine change without revealing whether it is an improvement or a downgrade.
This effect is not observed for variable machines that support human performance in difficult tasks, probably because the participant is not able to clearly identify that the machine is performing accurately at the beginning.

It is notable to observe the distinction between outcomes from the static 50\% accuracy machine versus those from machines with varying accuracy levels, increasing or decreasing. 
While all three machines maintain an average accuracy of 50\%, they diverge in how their errors are distributed.
Based on our findings, we can deduce that for challenging tasks, randomly distributed errors throughout the interaction promote error camouflage and thus increase the influence factor, compared to machines that accumulate errors at the beginning or at the end of the interaction flow, which have lower influence factors.

\section{Conclusions and Future Work}
\label{sec:conclusions}

%\todo{In this section you briefly restate the main elements from the section above, and present limitations and future work}
%Conclusions:
%1. The difficulty of the task as perceived by the human shapes the effectiveness of a highly accurate static machine.
%2. The error induction of an inaccurate static machine outperforms the success induction of an accurate static machine when the task is difficult.
%3. Algorithmic appreciation is potentially transformable into automation bias when the user initially sees that the machine succeeds at easy tasks. This can be avoided by notifying the human that the machine has undergone changes.

We noted that the difficulty of tasks shapes human-machine interactions in a variety of ways, even when the complexity levels are relatively similar. In our study, there is merely a 10\% difference in average human accuracy between \textit{simple} and \textit{complex} tasks.
Nevertheless, this small gap appears to be sufficient to notably change the influence of decision support. It is therefore crucial to understand that this challenge relates more to how the participant perceives the task than to their actual skill in solving it.
Thus, high difficulty can affect the effectiveness of an accurate machine and can enhance the influence of an inaccurate machine. 
Conversely, low difficulty can enhance automation bias in the case of variable machines, more specifically those machines that start out accurate (thus eliciting appreciation) and later deteriorate.

In this work we combined the interpretation of the influence factor with the confirmation factor, and that helped us to better understand patterns of interaction. However, the concept of influence allows for many approaches that need to be considered. We have noted that this influence can be negative (aversion), positive (appreciation/automation bias) or neutral (no influence). We observed a tendency towards no influence in easy tasks, and a more erratic tendency towards influence in the case of difficult tasks. This highlights the complexity of measuring influence, necessitating further study on when it is beneficial or detrimental for task completion.
We also observed that participants can tell accurate (95\%) from inaccurate (5\%) decision support when the task is easy, but lose this ability when the task is hard, which makes them more prone to be misled.

A significant constraint identified is the insufficient consideration of real-world, practical concerns. Facial recognition systems are used in complex ethical domains like immigration and law enforcement, where delegating decisions to machines can lead to anonymity, psychological detachment, and invisibility \cite{ostermaier2017spot,kobis2019intuitive,hancock2015deception}. These factors may inadvertently promote unethical actions. It is therefore urgent and necessary to extend research on hybrid systems and their corresponding interaction patterns into domains more closely linked to real-world application fields.

\smallskip\noindent
\textbf{Code and Data} are available in \href{https://github.com/ealmenzar/HumanResponse-DSS-FaceMatching}{\texttt{HumanResponse-DSS-FaceMatching}} repository.\footnote{https://github.com/ealmenzar/HumanResponse-DSS-FaceMatching}

\smallskip\noindent
\textbf{Acknowledgments}\; This work has been partially supported by:
the Department of Research and Universities of the Government of Catalonia (SGR 00930);
and the Maria de Maeztu Units of Excellence Programme CEX2021-001195-M, funded by MICIU/AEI/10.13039/501100011033.

\enlargethispage{\baselineskip}

%%%%%%%%% BIBLIOGRAPHY %%%%%%%%

\bibliographystyle{vancouver}
\bibliography{mybib}

%%%%%%%%% APPENDICES %%%%%%%%

\end{document}